\documentclass{article}
\usepackage{style}

\title{Quantum Computing, Ising formulation, and the Traveling Salesman Problem}

\author[1]{Omer Gurevich}
\author[1]{Maor Matityahu}
\author[1,2]{Tal Mor}

\affil[1]{Computer Science Department, Technion, Haifa, Israel\footnote{Email:  \texttt{omergu@campus.technion.ac.il}}}
\affil[2]{The Helen Diller Quantum Center, Technion, Haifa, Israel}

\date{\today}

\begin{document}

\maketitle
\begin{abstract}
    
Ising formulation is important for many NP problems (Lucas, 2014). 
This formulation enables
implementing novel quantum computing methods including Quantum Approximate Optimization Algorithm and Variational Quantum Eigensolver (VQE). Here, we investigate closely the traveling salesman problem (TSP).   

First, we present some non-trivial issues related to Ising model view versus a realistic salesman.

Then, focusing on VQE we discuss and clarify the use of: 
a.-- Conventional VQE and how it is relevant as a novel SAT-solver; 
b.-- Qubit efficiency and its importance in the Noisy Intermediate Scale Quantum-era; 
and 
c.-- the relevance
and importance of a novel approach named Discrete Quantum Exhaustive Search
(Alfassi, Meirom, and Mor, 2024), for enhancing VQE and other methods using mutually unbiased bases.

The approach we present here in details can potentially be extended 
for analyzing approximating and solving various other NP complete problems. Our approach can also be
extended beyond the Ising model and beyond the class NP, for example to the class Quantum Merlin Arthur (QMA) of problems, relevant for quantum chemistry and for general spin problems.

\end{abstract}
\section{Introduction}
\label{sef:Intro}
\subsection{Motivation}
\label{sef:Motivation}
Quantum computers promise to change the computer science and engineering world.
In practice however, there are three major limitations depending on the type
of problems:
A.-- For some problems it is fully believed (and in some case 
even proven under reasonable assumptions) that quantum computers will
exponentially outperform classical ones. However, it seems that the tiny needed
error-rate, the large number of qubits, and the need for fault-tolerance
quantum error correction, limit the applicability in those cases for a decade
or maybe even several decades in the future. The most typical and well known
examples for that type of problems are factoring large numbers \cite{ref:shor1999polynomial} and quantum
simulations for various condensed-matter problems \cite{ref:Lloyd1996simulations}. 
B.-- For some problems there is also a great reason to believe that there is
an exponential quantum advantage, possibly already in the near future (named
NISQ era, noisy intermediate scale quantum era), yet these problems
are man-made especially for the purpose of proving the quantum advantage, e.g., see  \cite{ref:Aaronson2011Boson, ref:GoogleSupremacy2019,ref:IBMRebuttal2019}. 
C.-- For many other problems \cite{ref:Bauer2020ChemistryAlgos,ref:Symons2023optimization} there is less theoretical ground (i.e., no proof
under any reasonable assumption) for an exponential quantum advantage, yet
these problems are of extreme importance, hence we expect a competition between
quantum heuristics and classical heuristics in the near future, i.e., in the
NISQ era.

Among the approaches relevant for the abovementioned type-C problems, the VQA \cite{ref:cerezo2021VQA} in
general and the VQE in particular, are the leading approaches. VQE is a great 
method for analyzing problems for which Ising model Hamiltonian or
Heisenberg model Hamiltonian can be written.

While it is of course not expected
that a quantum computer will solve any NP-complete problem, it is expected to
provide the analogue of the many SAT-solvers \cite{ref:Biere2009SAT}, that we can call LH-solvers,
since Local-Hamiltonian is the natural generalization of SAT when extending
from NP problems in the classical world to QMA  problems in the quantum world \cite{ref:Kitaev2002QMA}.

Similarly to the ability of SAT-solvers to solve SAT and many other NP hard and
NP complete problems efficiently, either by solving optimal approximations,
or by finding and solving easy sub-classes, we expect the same from LH-solvers.
Actually, since the quantum space is far beyond the classical one (e.g., for a
qubit versus a bit, it is the whole surface of a sphere instead of just two
points the north pole and the south pole), we expect to see in the future
novel quantum approaches to NP problems, novel quantum-inspired classical
approaches to NP problems, and novel quantum approaches to problems that are
harder than NP problems, such as finding the ground state energy of chemical 
molecules (e.g., Density Functional Theory (DFT) method is proven  to be QMA complete \cite{ref:Schuch2009DFT}). 

In the classical NP world, exhaustive search on small-size problems 
is majorly used when planning to find a SAT-solver (or such) for NP problems.
Motivated by that approach, discrete quantum exhaustive search (DQES) \cite{ref:Meirom2024MUBs} has
been recently suggested (by  Alfassi, Meirom, and Mor) and designed for potentially improving quantum
algorithms, and potentially solve two big problems of reaching a solution
via an iterative optimization process such as VQA: 
the problems of local minima and of barren plateaus \cite{ref:mcclean2018barren,ref:larocca2025barren}.

Our focus here is VQE, for Ising Hamiltonian. And more specifically 
we focus on the Traveling Salesman Problem (TSP),
an extremely important NP-complete problem. Although 
we focus here on the TSP, the methods we discuss and demonstrated 
are relevant far beyond the TSP.

\subsection{Hamiltonian Path and Hamiltonian Cycle}
\label{sec:HPHC}
\begin{figure}[H]
    \centering
    \begin{minipage}{0.35\textwidth}
        \centering
        \begin{tikzpicture}
            \tikzset{VertexStyle/.style={draw, circle, text=black, fill=white, inner sep=0.1cm}}
    
            \SetUpEdge[labelstyle={sloped}, labelcolor=black]
            \GraphInit[vstyle=Normal]
            \SetGraphUnit{3}
    
            \Vertex[x=0, y=2.5]{1}  
            \Vertex[x=2.5, y=2.5]{2}  
            \Vertex[x=0, y=0]{3}  
            \Vertex[x=2.5, y=0]{4}  
    
            \tikzset{EdgeStyle/.style={->, >=latex, color=blue, line width=2pt}}
            \Edge(2)(1)
            \Edge(1)(4)
            \Edge(4)(3)
    
            \tikzset{EdgeStyle/.style={->, >=latex, color=blue, line width=0.5pt}}
            \Edge(2)(3)
    \end{tikzpicture}
    \\
    \textbf{(a)}   
    \end{minipage}%
    \hspace{20pt}
    \begin{minipage}{0.35\textwidth}
        \centering
        \begin{tabular}{|c|c|c|c|c|}
            \hline
            \backslashbox[10mm]{\textbf{v}}{\textbf{t}} & \textbf{1} & \textbf{2} & \textbf{3} & \textbf{4} \\
            \hline
            \textbf{1} & 0 & 1 & 0 & 0 \\
            \hline
            \textbf{2} & 1 & 0 & 0 & 0 \\
            \hline
            \textbf{3} & 0 & 0 & 0 & 1 \\
            \hline
            \textbf{4} & 0 & 0 & 1 & 0 \\
            \hline
        \end{tabular}
        \\
        \vspace{19pt}
        \textbf{(b)}
    \end{minipage}
    \caption{(a) Example of a Hamiltonian path problem instance on a 4-node directed graph. The (unique) solution is in bold. (b) A valid $N \times N$ assignment table representation of the variables $x_{v,t}$}
    \label{fig:HPath}
\end{figure}
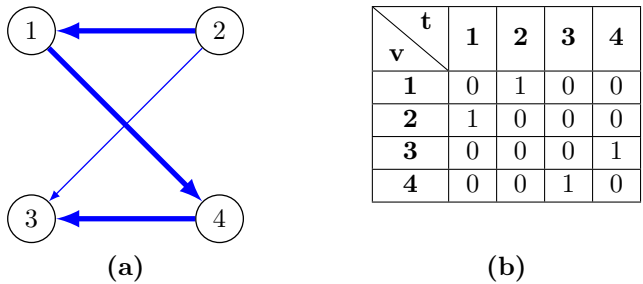

Given a graph $G=(E,V)$, directed or undirected, a  Hamiltonian path is a path that visits every node exactly once. Similarly, Hamiltonian cycle is the same where in addition the path ends at the starting point and returns to the starting node.
\\
Note that in the Hamiltonian cycle, we describe a solution by the order of nodes that define the cycle, and therefore any cyclic permutation is an equivalent solution. Anti-cyclic permutations are also equivalent solutions if the graph is undirected. In Hamiltonian path this symmetry does not appear. In what follows, we use a quantum algorithm to solve problems that are based on the Hamiltonian cycle. By the principles of quantum mechanics, a superposition of such equivalent solutions is also a valid solution, a phenomenon which has no equivalency in the classic world. By fixing a starting point, this degeneracy can be avoided. However, the two-fold degeneracy that appears due to direction in undirected graphs cannot be avoided.
\\
Determining whether a graph $G$ contains a Hamiltonian path (or a cycle) is an NP-complete problem. However, it can be solved in some cases, in many ways. \\
Here, we follow a solution suggested by Lucas~\cite{ref:Lucas_2014}. This solution also enables us to construct the path (or cycle) if it exists. Starting with the Hamiltonian path, we denote the vertices by $v=1,2,...,|V|$, and we denote $|V|\coloneqq N$. Next, we define binary variables $x_{v,t}$ that take the value $1$ if and only if node $v$ is the $t$\textsuperscript{th} to be visited in the cycle. A proposed solution can be thus represented with an $N \times N$ table that contains the assignments of the variable $x_{v,t}$. For an example of a Hamiltonian path problem instance and a valid solution, see \Fig{fig:HPath}.
\\

\begin{figure}[H]
    \centering
    \begin{minipage}{0.35\textwidth}
        \centering
        \begin{tikzpicture}
            \tikzset{VertexStyle/.style={draw, circle, text=black, fill=white, inner sep=0.1cm}}
    
            \SetUpEdge[labelstyle={sloped}, labelcolor=black]
            \GraphInit[vstyle=Normal]
            \SetGraphUnit{3}
    
            \Vertex[x=0, y=2.5]{1}  
            \Vertex[x=2.5, y=2.5]{2}  
            \Vertex[x=0, y=0]{3}  
            \Vertex[x=2.5, y=0]{4}  
    
            \tikzset{EdgeStyle/.style={->, >=latex, color=blue, line width=2pt}}
            \Edge(2)(1)
            \Edge(1)(4)
            \Edge(4)(3)
            \Edge(3)(2)

            \tikzset{EdgeStyle/.style={->, >=latex, color=blue, line width=0.5pt}}
            \Edge(2)(4)
            \Edge(3)(1)          
    \end{tikzpicture}
    \\
    \textbf{(a)}   
    \end{minipage}%
    \hspace{20pt}
    \begin{minipage}{0.35\textwidth}
        \centering
        \begin{tabular}{|c|c|c|c|c|c|}
            \hline
            \backslashbox[10mm]{\textbf{v}}{\textbf{t}} & \textbf{1} & \textbf{2} & \textbf{3} & \textbf{4} & 
            \textbf{5} \\
            \hline
            \textbf{1} & 0 & 1 & 0 & 0 & 0 \\
            \hline
            \textbf{2} & 1 & 0 & 0 & 0 & 1 \\
            \hline
            \textbf{3} & 0 & 0 & 0 & 1 & 0 \\
            \hline
            \textbf{4} & 0 & 0 & 1 & 0 & 0 \\
            \hline
        \end{tabular}
        \\
        \vspace{12pt}
        \textbf{(b)}
    \end{minipage}
    \caption{(a) Example of a Hamiltonian cycle problem instance on a 4-node directed graph. A solution (up to cyclic permutation) is in bold. (b) A valid $N \times (N+1)$ assignment table representation of the variables $x_{v,t}$}
    \label{fig:HCyc}
\end{figure}
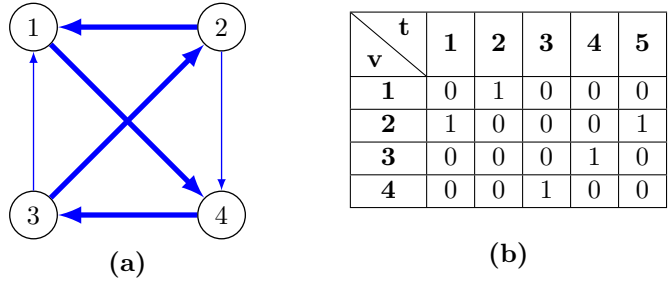

The way to formulate the Hamiltonian cycle problem is very similar to what we did for the Hamiltonian path. To close a cycle, one more node has to be visited. Then we shall use the same binary variables $x_{v,t}$, this time adding $N$ more variables for $x_{v,t=N+1}$. 
For an example of a Hamiltonian cycle problem instance and a valid solution, according to the proposed naive treatment, see \Fig{fig:HCyc}.
\\
As the problem deals with finding a cycle, it is enough to define the cycle with $N$ vertices $v_1,\ldots,v_N$, knowing that $v_{N+1}=v_1$. For a detailed discussion about this way of reducing the number of required variables, as well as other ways, see Secs. \ref{sec:Qubit-eff} and \ref{sec:Qubit-eff2}.
\\

To solve the Hamiltonian cycle, we begin with formulating the constraints that must hold in a valid solution, using the variables $x_{v,t}$, As was done in~\cite{ref:Lucas_2014}:
\begin{align}
    \label{eq:con1Lucas}
    & \sum_{t=1}^{N} x_{v,t}=1, \quad v=1,2,\ldots,N,
    \\
    \label{eq:con2Lucas}
    & \sum_{v=1}^{N} x_{v,t}=1, \quad t=1,2,\ldots,N,
    \\
    \label{eq:con3Lucas}
    & \sum_{(u,v)\notin E} x_{u,t} x_{v,t+1} = 0, \quad t=1,2,\ldots,T.
\end{align}
From \Eq{eq:con1Lucas} it is ensured that every node appears exactly once in the path (or cycle), where \Eq{eq:con2Lucas} states that for any $t=1,2,\ldots,N$ there must be a single $t$\textsuperscript{th} node in it. In addition, from \Eq{eq:con3Lucas} we get that the path (or cycle) is connected. From these expressions (\ref{eq:con1Lucas})-(\ref{eq:con3Lucas}) altogether, we see that the Hamiltonian cycle problem is equivalent to finding the values of $x_{vt}$ that minimize the energy of the Hamiltonian:
\begin{align}
\label{eq:HCyc}
    H_A
    =A\sum_{v=1}^{N} \left(1-\sum_{t=1}^{N} x_{v,t}\right)^2  
    +A\sum_{t=1}^{N} \left(1-\sum_{v=1}^{N} x_{v,t}\right)^2 
    +A \sum_{(uv)\notin E}\sum_{t=1}^{N}x_{u,t}x_{v,t+1}, 
\end{align}
for some constant $A>0$, where we identify $N+1$ with $1$, as we mentioned before. This identification allows for using $N$ bits less than without it. For the Hamiltonian path problem, the last sum should run up to $N-1$ instead of $N$. Any assignment of the variables $x_{v,t}$ for which $H_A=0$ encodes a valid Hamiltonian cycle. 

\subsection{The traveling salesman problem}
\label{sec:TSP}
\begin{figure}[H]
    \centering

    \begin{minipage}{0.35\textwidth}
        \centering
        \begin{tikzpicture}
        \tikzset{VertexStyle/.style={draw, circle, text=black, fill=white, inner sep=0.1cm}}
        
        \SetUpEdge[color=blue, labelcolor=white, labelstyle=sloped]
        \GraphInit[vstyle=Normal]
        \SetGraphUnit{2}

        \Vertex[x=0, y=0]{3} 
        \Vertex[x=4, y=0]{4}
        \Vertex[x=0, y=2.5]{1}
        \Vertex[x=2.5, y=2.5]{2}

        \tikzset{EdgeStyle/.style={-,>=latex}} 
        \Edge[lw=1pt, label=\scriptsize{$1$}](1)(2)
        \Edge[lw=2.5pt, label=\scriptsize{$3$}](2)(3)
        \Edge[lw=1pt, label=\scriptsize{$8$}](3)(4)
        \Edge[lw=2.5pt, label=\scriptsize{$5$}](4)(1)
        \Edge[lw=2.5pt, label=\scriptsize{$1$}](1)(3)
        \Edge[lw=2.5pt, label=\scriptsize{$2$}](2)(4)
    \end{tikzpicture} 
        \vspace{1em}
        \textbf{(a)}
    \end{minipage}%
    \hspace{40pt}
    \begin{minipage}{0.35\textwidth}
        \centering
                \begin{tabular}{|c|c|c|c|c|}
            \hline           \backslashbox[10mm]{\textbf{v}}{\textbf{t}} & \textbf{1} & \textbf{2} & \textbf{3} & \textbf{4}  \\
            \hline
            \textbf{1} & 0 & 1 & 0 & 0  \\
            \hline
            \textbf{2} & 0 & 0 & 0 & 1  \\
            \hline
            \textbf{3} & 1 & 0 & 0 & 0  \\
            \hline
            \textbf{4} & 0 & 0 & 1 & 0  \\
            \hline
        \end{tabular}
        \\
        \vspace{1em}
        \textbf{(b)}
    \end{minipage}
    \caption{(a) Example of a TSP instance on a 4-node undirected graph. The optimal solution (up to cyclic permutations, as well as two possible directions) is in bold (b) A valid 
    efficient (i.e., $N \times N$) assignment table representation of the variables $x_{v,t}$}
    \label{fig:TSP_NN}
\end{figure}
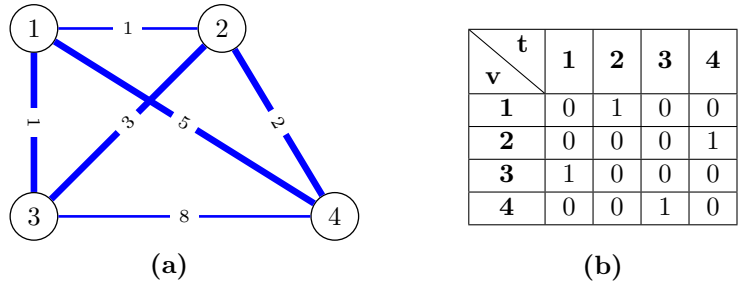

In the Travel Salesman Problem (TSP), a salesperson has to visit other cities and return to its city of origin, where every city is visited exactly once, and the distance of the total travel is minimized. 
\\
We can model the problem with a weighted graph $G=(E,V)$ (directed or undirected), where every edge $(u,v)\in E$ is assigned a weight (we call it cost) denoted by $c_{uv}\ge 0$.   
One invalid solution allows, naively, the salesman to fail  in his last move, and choose an edge that is cheaper than the one that closes the cycle. Our formulation does not allow this failure, by relying on the Hamiltonian cycle formulation. 
\\
Furthermore, in order to slightly decrease the complexity of the problem - we can reduce using some bits (in the table) for the last time step, by employing the cyclic properties of the problem, and specifically, by identifying the initial point with the last point of the cycle while not writing it explicitly in a table.
 In \Fig{fig:TSP_NN}, we show a TSP instance and its solution, with respect to this efficient formulation
As we said, the salesman must traverse a Hamiltonian cycle on the graph, such that the joint cost of the edges in the cycle is minimized. Therefore, we can extend our formulation for the Hamiltonian cycle stated in the previous section, by including this minimization requirement, and obtain an encoding of the problem via a Hamiltonian cost function. To achieve that, we add the term 
\begin{align}
\label{eq:HB}
    H_B = B \sum_{(uv)\in E} c_{uv} \sum_{t=1}^{N}x_{u,t}x_{v,t+1},
\end{align}
to the Hamiltonian. The total Hamiltonian that encodes the TSP is therefore given by
\begin{align}
\label{eq:HTSP-simple}
H_{TSP} = H_A + H_B,
\end{align} with  $H_A$ from \Eq{eq:HCyc}, where the constants $A,B$ must be chosen such that penalties for bad solutions are high enough, and only valid solutions are energetically preferable. Further details on how these constants are chosen are given for general graphs in \Sec{sec:AB-General}, and for the complete graphs in \Sec{sec:AB-Comp}. 
\section{A Note about Penalties in General Graphs}
\label{sec:AB-General}
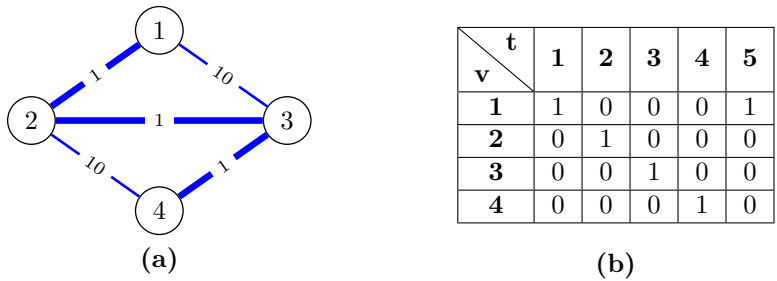
\begin{figure}[H]
    \centering
    \begin{minipage}{0.35\textwidth}
        \centering
        \begin{tikzpicture}
        \tikzset{VertexStyle/.style={draw, circle, text=black, fill=white, inner sep=0.1cm}}
        
        \SetUpEdge[color=blue, labelcolor=white, labelstyle=sloped]
        \GraphInit[vstyle=Normal]
        \SetGraphUnit{2}

        \Vertex[x=0, y=0]{1} 
        \Vertex[x=-1.7, y=-1.2]{2}
        \Vertex[x=1.7, y=-1.2]{3}
        \Vertex[x=0, y=-2.4]{4}

        \tikzset{EdgeStyle/.style={-,>=latex}} 
        \Edge[lw=2.5pt, label=\scriptsize{$1$}](1)(2)
        \Edge[lw=2.5pt, label=\scriptsize{$1$}](2)(3)
        \Edge[lw=2.5pt, label=\scriptsize{$1$}](3)(4)
        \Edge[lw=1pt, label=\scriptsize{$10$}](1)(3)
        \Edge[lw=1pt, label=\scriptsize{$10$}](2)(4)
    \end{tikzpicture} 
        \vspace{1em}
        \textbf{(a)}
    \end{minipage}%
    \hspace{40pt}
    \begin{minipage}{0.4\textwidth}
        \centering
                \begin{tabular}{|c|c|c|c|c|c|}
            \hline           \backslashbox[10mm]{\textbf{v}}{\textbf{t}} & \textbf{1} & \textbf{2} & \textbf{3} & \textbf{4} & \textbf{5} \\
            \hline
            \textbf{1} & 1 & 0 & 0 & 0 & 1  \\
            \hline
            \textbf{2} & 0 & 1 & 0 & 0 & 0  \\
            \hline
            \textbf{3} & 0 & 0 & 1 & 0 & 0  \\
            \hline
            \textbf{4} & 0 & 0 & 0 & 1 & 0  \\
            \hline
        \end{tabular}
        \\
        \vspace{1em}
        \textbf{(b)}
    \end{minipage}
    \caption{A counter example to correctness of \cite{ref:Lucas_2014} assuming $A=11,B=1$. \: For an efficient formulation, (a) Undirected graph for the TSP instance with optimal solution in bold. By definition, there is also a step from node 5 back to node 1, even though there is no edge joining these two nodes. The only legitimate solutions with node 1 as the starting point are the ones with a cost of 22. (b) 
    A non-efficient yet easy to understand (i.e., $N \times (N+1)$) assignment table representation for an invalid, yet optimal solution, where a single penalty is paid for the missing edge.}
    \label{fig:counterEx}
\end{figure}

In \cite{ref:Lucas_2014} it was suggested that the condition on $A,B$ from \Eq{eq:HTSP-simple} could be
\begin{equation}
\label{eq:AB-ineq}
    0<B\max(c_{u,v})<A.   
\end{equation}
However, we noted that this condition is not sufficient in general. 
In \Fig{fig:counterEx} we present a counter-example graph, along with its corresponding table assignment of an invalid, yet lowest-cost solution.  
 
Suppose that $B=1$ and $A=11>B\max(c_{u,v})$. In the counter-example, we have ground-energy $H_{TSP}=14$, whereas the valid solutions are given by the cycle $1-2-4-3-1$ (or the opposite cycle), and they cost $H_{TSP}=22$.

A safer condition than the one given in \Eq{eq:AB-ineq} is $0<N\cdot B \max(c_{u,v})<A$. For any $A,B>0$ satisfying this, the optimal solution is guaranteed to be valid. 

It may well be that for a complete graph, the condition in \cite{ref:Lucas_2014} is sufficient. An intuitive explanation is given in the next subsection.

We note that the Hamiltonian provided in~\cite{ref:Lucas_2014} limits the free will of the salesman. For example, the salesman cannot perform more than N steps, and if performing exactly N steps than his last step must be back to the starting point.

\subsection{Penalties in Complete Graphs}
\label{sec:AB-Comp}
In TSP, it is common to assume a complete graph \cite{ref:Lucas_2014}. We now show that under this assumption, for any choice of $A,B>0$ such that \Eq{eq:AB-ineq} holds, we are promised that the optimal valid tour indeed minimizes the energy of $H_{TSP}$. Using the table representation of the assignments leads to an intuitive proof of this fact. The logic is as follows. 

In the assignment table, a cell corresponding to some variable $x_{v,t}$ is called \textit{empty} if $x_{v,t}=0$ and \textit{occupied} otherwise. According to Eqs. (\ref{eq:con1Lucas}) and (\ref{eq:con2Lucas}), we note that the table of any invalid solution has at least one empty row (or column), or at least one row (or column) with more than one cell occupied. On the other hand, a table with exactly one occupied cell at each row (and therefore also at each column) describes a valid solution. This is due to the fact that all edges exist, so no penalty is caused from \Eq{eq:con2Lucas} (Alternatively, this constraint could be ignored for a complete graph in the first place).

Assume by contradiction that some invalid solution $x'$ has minimal energy. We can turn it into a valid solution in a way that reduces its energy, arriving at a contradiction. There are many ways to do this. For example, we can go row by row, and when we encounter a row with $k+1$ full cells, for $k>0$, We can empty $k$ of these cells, arbitrarily. We may have removed some edges in the process, which lowers the contribution from $H_B$. Moreover, now a penalty of $k^2A$ due to the over-occupied row is removed, where new penalties may appear due to columns that are maybe empty. However, these new penalties are no greater than $k^2A$. So the energy was not increased. Continuing in this fashion for all rows, and then for all columns, we arrive at a table with columns and rows that are either empty or with single-occupation, without increasing the energy. The number of empty rows must be the number of
empty columns. 

Next, we pick an empty row and fill one of the cells that also belongs to an empty column. We repeat this until exactly each cell is occupied in each row and in each column. At each iteration, two edges $e,f$ are added to the solution, and two violations are corrected. Hence, $H_A$ is reduced by $2A\coloneqq \Delta H_A$ where $H_B$ is increased by $B(c(e)+c(f))\coloneqq \Delta H_B$. By \Eq{eq:AB-ineq}, at each iteration, we only reduced the energy, since $\Delta H_A+\Delta H_B < 0$. 

Note that the process ended with a valid tour, and the cost was lowered, where the optimal cycle costs even less. We conclude that the condition in \Eq{eq:AB-ineq} for the penalty coefficient $A$ is a good choice for complete graphs, where any $B>0$ works. 

\subsection{A Note about Realistic Salesman}

We note that the Hamiltonian provided in~\cite{ref:Lucas_2014} and copied in the introduction here, limits the free will of the salesman. For example, the salesman cannot perform more than N steps, and once  performing exactly N steps than his last step must be back to the starting point.

On the other hand, the Hamiltonian formulation allows various things that a realistic salesman cannot do. First, a realistic salesman cannot travel along non-existing path. We suggest to resolve this problem by using a much higher penalty for the third term in the Hamiltonian for the Hamiltonian-cycle. Second, a realistic salesman cannot be in two nodes at the same time. This problem again can be resolved by a much  higher penalty relative to the penalty of size $A$ for the realistic salesman returning twice to the same node to save some cost. A last improvement can be to allow a realistic salesman to perform as many steps as he like (more than $N$, less than $N$, and add a penalty that grows as the difference from $N$ grows.

Of course any such improvement of the Ising Hamiltonian will add a complexity, so we stick to the Hamiltonian-path based Hamiltonian as Lucas did.
\section{A Note about Qubit Efficiency}
\label{sec:Qubit-eff}
Saving qubits requires some caution. In Table \ref{tab:valid-invalid}, we show the problem that could result from an efficient representation, by looking at two different solutions, a correct one and a failed one, when described by completing the partial table into an $N \times (N+1)$ table, for that same undirected graph. 
\begin{table}[H]
    \centering
    \begin{minipage}{0.35\textwidth}
        \centering
                \begin{tabular}{|c|c|c|c|c|c|}
            \hline
            \backslashbox[10mm]{\textbf{v}}{\textbf{t}} & \textbf{1} & \textbf{2} & \textbf{3} & \textbf{4} & 
            \textbf{5} \\
            \hline
            \textbf{1} & 0 & 1 & 0 & 0 & 0 \\
            \hline
            \textbf{2} & 0 & 0 & 0 & 1 & 0 \\
            \hline
            \textbf{3} & 1 & 0 & 0 & 0 & 1 \\
            \hline
            \textbf{4} & 0 & 0 & 1 & 0 & 0 \\
            \hline
        \end{tabular}
        \vspace{1em}
        \textbf{(a)}
    \end{minipage}
    \hspace{20pt} 
    \begin{minipage}{0.35\textwidth}
        \centering
                \begin{tabular}{|c|c|c|c|c|c|}
            \hline
            \backslashbox[10mm]{\textbf{v}}{\textbf{t}} & \textbf{1} & \textbf{2} & \textbf{3} & \textbf{4} & 
            \textbf{5} \\
            \hline
            \textbf{1} & 0 & 1 & 0 & 0 & 1 \\
            \hline
            \textbf{2} & 0 & 0 & 0 & 1 & 0 \\
            \hline
            \textbf{3} & 1 & 0 & 0 & 0 & 0 \\
            \hline
            \textbf{4} & 0 & 0 & 1 & 0 & 0 \\
            \hline
        \end{tabular}
        \vspace{1em}
        \textbf{(b)}
    \end{minipage}

    \caption{(a) A valid TSP assignment solution described by the full table. (b) An invalid TSP assignment for the same instance, demonstrating violation of constraints via the full table. The demonstration of an efficient yet invalid solution is relevant also for the Hamiltonian cycle problem.}
    \label{tab:valid-invalid}
\end{table}
Another invalid situation due to using the efficient representation could be left unnoticed is in case there is no edge connecting node 2 (visited at time step 4) and node 3 (visited at time step 5). 
\section{Fixing a Starting Point}
\label{sec:FixStartPt}
As mentioned in \Sec{sec:HPHC}, in Hamiltonian cycle, every Hamiltonian cycle can be expressed in $N$ different ways for directed graphs, and $2N$ ways for undirected graphs. If we add the requirement that the salesman starts from a specific node, say $v=1$, then this symmetry is broken (there can still be two equivalent cycles for undirected graphs, but this is a $O(1)$ redundancy). \\
Therefore, we add the constraint $x_{1,1}=1$,  that forces the salesman to start at $v=1$. The Hamiltonian is updated accordingly:
\begin{equation}
    \begin{split}
\label{eq:HTSP}
    H_{TSP}
    &=A(1-x_{1,1})^2+  A\sum_{v=1}^{N} \left(1-\sum_{t=1}^{N} x_{v,t}\right)^2  
    +A\sum_{t=1}^{N} \left(1-\sum_{v=1}^{N} x_{v,t}\right)^2 
    \\
    &+A \sum_{(uv)\notin E}\sum_{t=1}^{N}x_{u,t}x_{v,t+1}
    +B \sum_{(uv)\in E} c_{uv} \sum_{t=1}^{N}x_{u,t}x_{v,t+1}, 
\end{split}
\end{equation}
Note that the variable $x_{1,1}$ is now actually a constant by the constraints. This hints that we can encode the problem with less variables, as we will see in \Sec{sec:Qubit-eff2}. This was also implied from the redundancy in solutions that we mentioned.  
\section{Ising Form of the Hamiltonian}
\label{sec:IsingForm}
Before we are able to run a VQA to solve the TSP, we need to rewrite the Hamiltonian from \Eq{eq:HTSP} in Ising form.
To achieve this, we express the Hamiltonian as a function of the spin variables $s_{v,t}=\pm1$ corresponding to eigenvalues of eigenstates $\ket{x_{v,t}}$ of the Pauli-$Z$ operator, rather than as a function of the binary variables.\\
We obtain this by substituting $x_{u,v} \mapsto \frac{1-s_{u,v}}{2}$ in \Eq{eq:HTSP}. Using the fact that for $x^2=x$ for a binary variable $x$, we can calculate the different terms in the Hamiltonian with the spin variables.  
After gathering terms and ignoring the constant part which is irrelevant for optimization, we obtain the Ising Hamiltonian:
\begin{equation}
\label{eq:HTSP_Ising}
\begin{split}
     H_{TSP}^{\text{(Ising)}} &= A(1 - N) \sum_{v=1}^{N} \sum_{t=1}^{N} s_{v,t} \\
&\quad    + \frac{A}{2} \left( s_{1,1} + \sum_{t=1}^{N} \sum_{1 \leq u < v \leq N} s_{u,t} s_{v,t} 
    + \sum_{v=1}^{N} \sum_{1 \leq s < t \leq N} s_{v,s} s_{v,t} \right) \\
    &\quad + A \sum_{(u,v) \notin E} \sum_{t=1}^{N} 
    \left( s_{u,t} s_{v,t+1} - s_{u,t} - s_{v,t+1} \right) \\
    &\quad + B \sum_{(u,v) \in E} c_{u,v} \sum_{t=1}^{N} 
    \left( s_{u,t} s_{v,t+1} - s_{u,t} - s_{v,t+1} \right).
\end{split}
\end{equation}
Note that we could find the constant part, at the cost of complicating the calculation. By having it, the ground state energy is exactly the optimal total cost. Nevertheless, the groundstate remains the same even if we drop it, and reconstructing the total cost from it is trivial.

\section{A Second Note about Qubit Efficiency}
\label{sec:Qubit-eff2}
We have already discussed how a formulation of the TSP can be wasteful in the number of qubits. In the naive approach, solving the Hamiltonian cycle problem requires $N\times(N+1)$
qubits, and so does the TSP. However, since we \textit{know} the solution is a cycle, we can identify $N+1\equiv 1$, which allows us to omit the spins
$s_{v,N+1}$.
This identification reduces the number of qubits by $N$, and leads to the Hamiltonian described 
in \Eq{eq:HTSP-simple}. \\
Furthermore, we presented another Hamiltonian in \Eq{eq:HTSP} that encodes the same problem but with an additional constraint; the first node in the cycle must be $v=1$.
The set of valid solutions remains the same, since the cycle is unchanged up to rotation. However, fixing the starting point enables an additional reduction in the number of qubits. In the TSP, the starting point is never visited at an intermediate step, so we can eliminate all variables corresponding to such visits. Hence, we are now able to reduce the number of qubits from $N^2$ to $(N+1)^2$, since we $\textit{know}$ the values some variables must take (in other words, they are constants):
\begin{equation}
\label{eq:trivial0}
       x_{1,t}=
       \begin{cases}
           1 & t=1 \\
           0 & t\neq 1        
       \end{cases},
\end{equation}
In Table (\ref{tab:efficientTSP}) we show which qubits take constant values in the TSP and, therefore, can be ignored, once we fix node $1$ as the starting point. This fixation of the starting point is then also demonstrated in the graph, see  \Fig{fig:starting-pt}, where we also present the final (efficient) table.

\begin{table}[h]
    \centering
    \renewcommand{\arraystretch}{1.5}
    \begin{tabular}{|c|c|c|c|c|c|}
        \hline
        \backslashbox[10mm]{\textbf{v}}{\textbf{t}} & \textbf{1} & \textbf{2} & \textbf{3} &  \textbf{4} &
        \textbf{5} \\
        \hline
        \textbf{1} & \cellcolor{red!25} 1 & \cellcolor{red!25}0 & \cellcolor{red!25}0 & \cellcolor{red!25}0 &
        \cellcolor{red!25}1\\
        \hline
        \textbf{2} & \cellcolor{red!25}0 & \cellcolor{green!25}0 & \cellcolor{green!25}1 & \cellcolor{green!25}0 &
        \cellcolor{red!25}0\\
        \hline
        \textbf{3} & \cellcolor{red!25}0 & \cellcolor{green!25}0 &\cellcolor{green!25}0 &\cellcolor{green!25}1 &
        \cellcolor{red!25}0\\
        \hline
        \textbf{4} & \cellcolor{red!25} 0 & \cellcolor{green!25}1 & \cellcolor{green!25}0 & \cellcolor{green!25}0 &
        \cellcolor{red!25}0\\
        \hline
    \end{tabular}
    \caption{Table representation of variables in $H_{TSP}$ from \Eq{eq:HTSP}. Known variables, with their values specified, are marked in red. The rest of the variables in general are unknown and are marked in green. Here they are taken to match the example in \Fig{fig:TSP_NN}, with the fixed starting point as shown in   \Fig{fig:starting-pt}.} 
    \label{tab:efficientTSP}
\end{table}
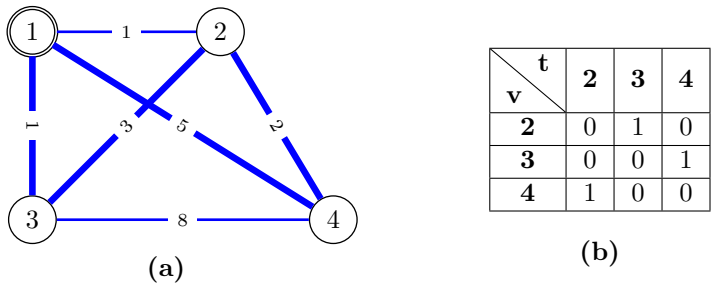
\begin{figure}[H]
    \centering

    \begin{minipage}{0.35\textwidth}
        \centering
        \begin{tikzpicture}
        \tikzset{VertexStyle/.style={draw, circle, text=black, fill=white, inner sep=0.1cm}}
        
        \SetUpEdge[color=blue, labelcolor=white, labelstyle=sloped]
        \GraphInit[vstyle=Normal]
        \SetGraphUnit{2}

        \Vertex[x=0, y=0]{3} 
        \Vertex[x=4, y=0]{4}
        \Vertex[style={draw, double, circle, fill=white},x=0, y=2.5]{1}
        \Vertex[x=2.5, y=2.5]{2}

        \tikzset{EdgeStyle/.style={-,>=latex}} 
        \Edge[lw=1pt, label=\scriptsize{$1$}](1)(2)
        \Edge[lw=2.5pt, label=\scriptsize{$3$}](2)(3)
        \Edge[lw=1pt, label=\scriptsize{$8$}](3)(4)
        \Edge[lw=2.5pt, label=\scriptsize{$5$}](4)(1)
        \Edge[lw=2.5pt, label=\scriptsize{$1$}](1)(3)
        \Edge[lw=2.5pt, label=\scriptsize{$2$}](2)(4)
    \end{tikzpicture} 
        \vspace{1em}
        \textbf{(a)}
    \end{minipage}%
    \hspace{40pt}
    \begin{minipage}{0.35\textwidth}
        \centering
                \begin{tabular}{|c|c|c|c|}
            \hline
            \backslashbox[10mm]{\textbf{v}}{\textbf{t}} & \textbf{2} & \textbf{3} & \textbf{4}  \\
            \hline
            \textbf{2} & 0 & 1 & 0  \\
            \hline
            \textbf{3} & 0 & 0 & 1  \\
            \hline
            \textbf{4} & 1 & 0 & 0  \\
            \hline
        \end{tabular}
        \\
        \vspace{1em}
        \textbf{(b)}
    \end{minipage}
    \caption{Reduced-qubit representation of after fixing node 1 as the starting point. One  solution out of two (namely, up to symmetry of the possible directions from node 1) is in bold.}
    \label{fig:starting-pt}
\end{figure}
To obtain a qubit efficient Hamiltonian $H_{TSP}^{\text{efficient}}$ in binary form, we plug \Eq{eq:trivial0} into \Eq{eq:HTSP}. Assuming undirected graph \footnote{The expression for directed graph does not differ by much. We show the undirected case for simplicity. Our simulated instances were on undirected graphs.}, this yields:
\\
\begin{equation}
\label{eq:redHTSP}
    \begin{split}
            & H_{TSP}^{\text{efficient}}  = \\
             & \quad A\left[ \sum_{v=2}^{N} \left( 1 - \sum_{t=2}^{N} x_{v,t} \right)^2 
             + \sum_{t=2}^{N} \left( 1 - \sum_{v=2}^{N} x_{v,t} \right)^2  \right] \\
            &\quad + A 
            \left[\sum_{(u,v) \notin E} \sum_{t=2}^{N-1} x_{u,t}x_{v,t+1}+
            \sum_{(1,v)\notin E} \sum_{t=2}^{N-1} (x_{v,2}
            + x_{u,N})
            \right]\\
            &\quad + B 
            \left[  \sum_{(u,v) \in E} 
            c_{u,v} \sum_{t=2}^{N-1} x_{u,t} x_{v,t+1} 
            +\sum_{(1,v)\in E} c_{1,v}( x_{v,2}+ x_{v,N})
            \right].
    \end{split}
\end{equation}
We also expressed this Hamiltonian in its Ising form, using the same transformation described in \Sec{sec:IsingForm}. This Ising Hamiltonian is the one used in the numerical experiments of \Sec{sec:exp}. However, the resulting expression is too lengthy and provides no additional insight, so we omit it.
\\

Overall, we were able to reduce the number of qubits from $N\times(N+1)$ to $(N-1)^2$. While this approach does not significantly reduce the space complexity—which remains $O(N^2)$—the reduction in the number of bits (or qubits, in the context of a quantum algorithm) can still be meaningful for VQAs in the NISQ era. Today’s quantum computers have a limited number of qubits, so even a small reduction in qubit requirements may be critical for enabling algorithms to run on non-trivial instances. 
\\
This efficiency in number of variables is also important for classical simulation of quantum algorithms. In the next section, for example, we simulate a TSP instance with 4 cities. Reducing the number of required variables from $4(4+1)=20$ to $(4-1)^2=9$ significantly reduces the running-time, since it scales exponentially with the number of variables.
\section{Numerical Experiments}
\label{sec:exp}
\subsection{DQES Landscape}
\label{sec:DQES}
We now present numerical results for calculating the TSP Hamiltonian energies for unique assignments $\{x_{v,t}\}$. These assignments correspond to the initial states used in the partial-DQES approach, presented in \cite{ref:Meirom2024MUBs}. There it was tested on optimization problems such as the MaxCut and transverse Ising. In partial-DQES, the Ising Hamiltonian energy is calculated for all the following states: For all choices of picking up $3$ qubits, we take all mutually unbiased bases (MUBs) elements on these $3$-qubits, where the rest of the qubits are at the zero state. There are $9$ such bases with $8$ basis elements in each. For more details on MUBs, and their importance in quantum information theory, see for example \cite{ref:cerf2002MUBs,ref:durt2005MUBs,ref:durt2010MUBs,ref:bruss1998MUBs,ref:lawrence2002MUBs,ref:vitoria2008MUBs,ref:butterley2007MUBs,ref:bandyopadhyay2002MUBs,ref:yu2023MUBsCirc}. So, for a Hamiltonian on $n$ qubits, we calculate the energy for ${n \choose 3}\cdot 72$ states. 

We now discuss a TSP instance for which we applied the method discussed above, which we refer to as \textit{energy landscape calculation}. In our simulations, we had to find the eigenvalues of the Ising Hamiltonian of interest. We mapped the expressions from \Eq{eq:redHTSP}, and used their Ising form for the TSP. The mapping was performed in a similar fashion to what we did in \Sec{sec:IsingForm} to derive $H_{TSP}$ in its Ising form (\Eq{eq:HTSP_Ising}).

We observe that, in general, we should not expect that one of the MUBs states is the eigenstate that corresponds to the optimal solution. However, as our experiments require $3$ time-steps only, we expect exactly $3$ qubits in the optimal solution to have value of one, where the rest should be zero. As $\ket{111}$ is an MUB state on $3$ qubits, we will obtain the optimal solution for one of the states we simulate.

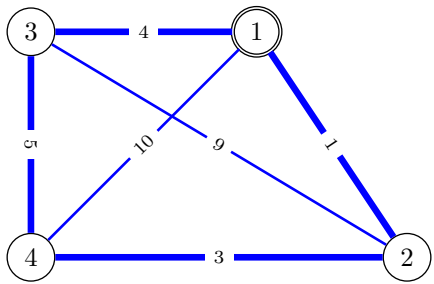
\begin{figure}[H]
    \centering
    \begin{tikzpicture}
        \tikzset{VertexStyle/.style={draw, circle, text=black, fill=white, inner sep=0.1cm}}
        
        \SetUpEdge[color=blue, labelcolor=white, labelstyle=sloped]
        \GraphInit[vstyle=Normal]
        \SetGraphUnit{2}

        \Vertex[x=0, y=0]{4} 
        \Vertex[x=5, y=0]{2}
        \Vertex[x=0, y=3]{3}
        \Vertex[style={draw, double, circle, fill=white}, x=3, y=3]{1}

        \tikzset{EdgeStyle/.style={-,>=latex}} 
        \Edge[lw=2.5pt, label=\scriptsize{$1$}](1)(2)
        \Edge[lw=1pt, label=\scriptsize{$9$}](2)(3)
        \Edge[lw=2.5pt, label=\scriptsize{$5$}](3)(4)
        \Edge[lw=1pt, label=\scriptsize{$10$}](4)(1)
        \Edge[lw=2.5pt, label=\scriptsize{$4$}](1)(3)
        \Edge[lw=2.5pt, label=\scriptsize{$3$}](2)(4)
    \end{tikzpicture} 
    \caption{An undirected graph defining a TSP and a instance for our simulations. One solution out of two (namely, up to symmetry of the possible directions from node 1) is in bold. Freedom up to cyclic permutation is removed after fixing node $1$ as the starting point.}
    \label{fig:Graph_K1}
\end{figure}

\begin{figure}[h!]
    \centering
    \includegraphics[width=0.8\linewidth]{{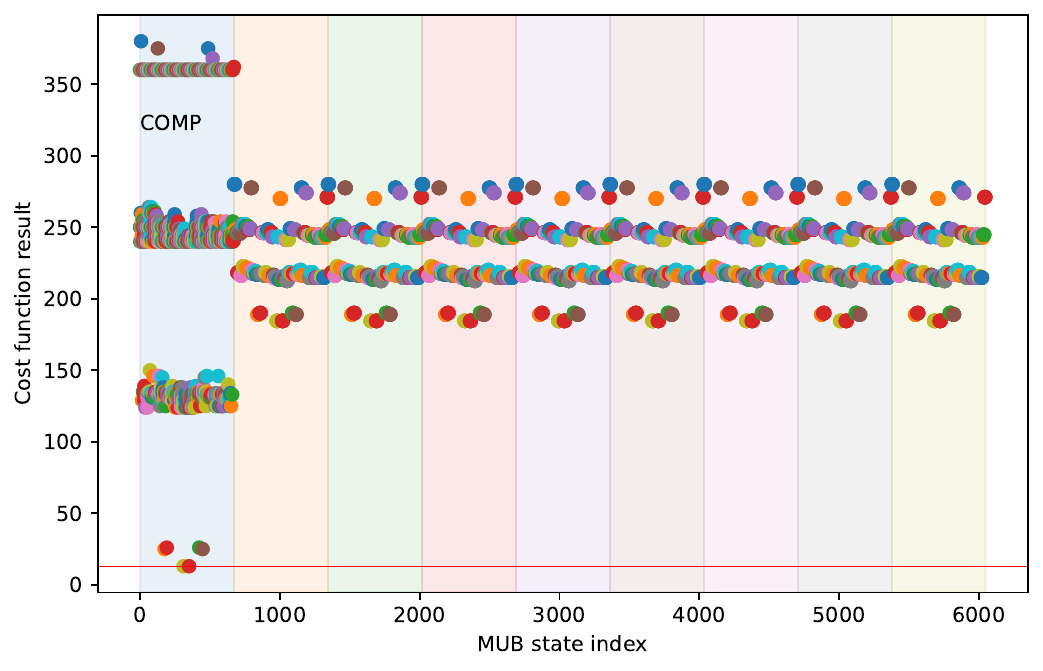}}
    \caption{Energy landscape of all partial-DQES initial states on two qubits, for the 4-node TSP instance of \Fig{fig:Graph_K1}. Two degenerate solutions with minimal energy corresponding to the optimal TSP cycles are apparent.}
    \label{fig:Landscape}
\end{figure}

We calculated the DQES landscape for a TSP instance with $N=4$ nodes, depicted in \Fig{fig:Graph_K1}.
As was demonstrated in Table (\ref{tab:efficientTSP}), the space efficient Hamiltonian which we use, is defined over $(N-1)^2=9$ qubits. We calculate the energy of all initial states of the partial-DQES method on $3$ qubits. In \Fig{fig:Landscape} we can see the energy for all such states.
Two states yield the minimum energy, which corresponds to the optimal TSP solution. These solutions are shown in Table (\ref{tab:Graph1Sol}), and are given by the cycles $1-2-4-3-1$ and $1-3-4-2-1$. Denoting the total cost of the travel by $C$, these two cycles have the same minimal cost $C=13$. Note that they correspond to the same cycle, but with opposite directions. This degeneracy is expected for undirected graphs, as discussed in \Sec{sec:HPHC}.
\begin{table}[H]
    \centering
    \begin{minipage}{0.35\textwidth}
        \centering
        \begin{tabular}{|c|c|c|c|c|c|}
            \hline    \backslashbox[10mm]{\textbf{v}}{\textbf{t}} & \textbf{1} & \textbf{2} & \textbf{3} & \textbf{4} & 
            \textbf{5} \\
            \hline
            \textbf{1} & 1 & 0 & 0 & 0 & 1 \\
            \hline
            \textbf{2} & 0 & 1 & 0 & 0 & 0 \\
            \hline
            \textbf{3} & 0 & 0 & 0 & 1 & 0 \\
            \hline
            \textbf{4} & 0 & 0 & 1 & 0 & 0 \\
            \hline
        \end{tabular}
        \\
        \vspace{12pt}
    \end{minipage}%
    \hspace{20pt}
    \begin{minipage}{0.35\textwidth}
        \centering
        \begin{tabular}{|c|c|c|c|c|c|}
            \hline
            \backslashbox[10mm]{\textbf{v}}{\textbf{t}} & \textbf{1} & \textbf{2} & \textbf{3} & \textbf{4} & 
            \textbf{5} \\
            \hline
            \textbf{1} & 1 & 0 & 0 & 0 & 1 \\
            \hline
            \textbf{2} & 0 & 0 & 0 & 1 & 0 \\
            \hline
            \textbf{3} & 0 & 1 & 0 & 0 & 0 \\
            \hline
            \textbf{4} & 0 & 0 & 1 & 0 & 0 \\
            \hline
        \end{tabular}
        \\
        \vspace{12pt}
    \end{minipage}
    \caption{The two optimal solutions of the TSP instance of \Fig{fig:Graph_K1}.}
    \label{tab:Graph1Sol}
\end{table}
\subsection{VQE Experiment on Top of DQES}
\label{sec:VQE}
The full scheme of DQES, as suggested in \cite{ref:Meirom2024MUBs}, involves running the VQE algorithm, starting from all MUB states (up to rotation), or all MUB states on $k$ out of $n$ qubits (with the other qubits potentially taken as $\ket 0$), in the case of partial-DQES. 
The idea behind DQES is that MUB states are evenly-spread in the Hilbert space, and therefore initializing VQE from such points can help bypass local minima and barren plateaus, which are the main challenges in VQAs. 

\begin{figure}[H]
    \centering

    \begin{minipage}{0.45\textwidth}
        \centering
        \includegraphics[width=\linewidth]{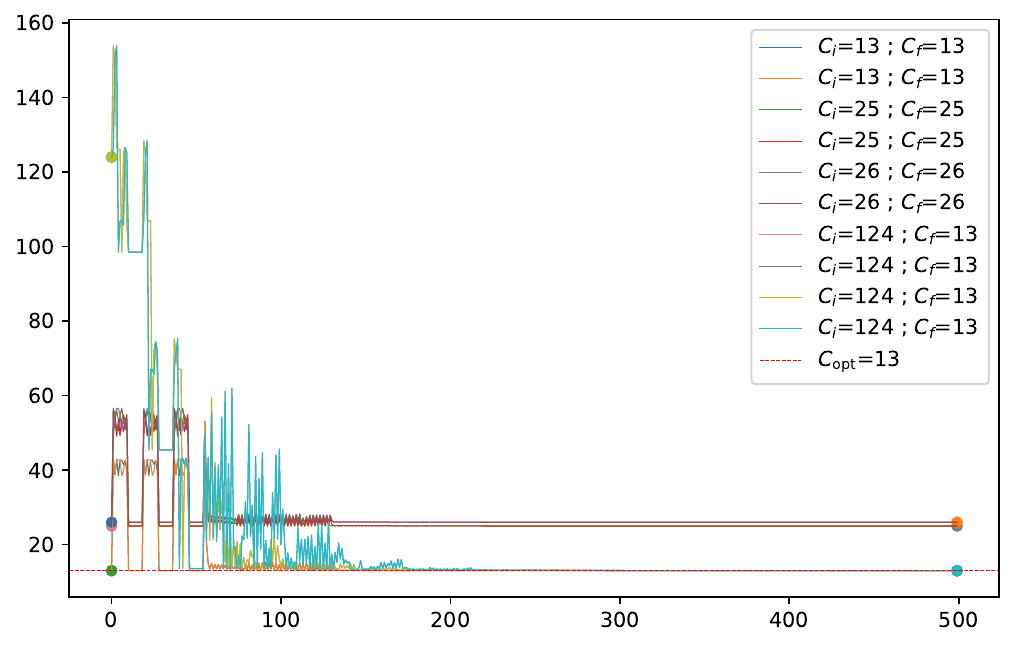}
        \vspace{0.5em}
        \textbf{(a)}
    \end{minipage}%
    \hfill
    \begin{minipage}{0.45\textwidth}
        \centering
        \includegraphics[width=\linewidth]{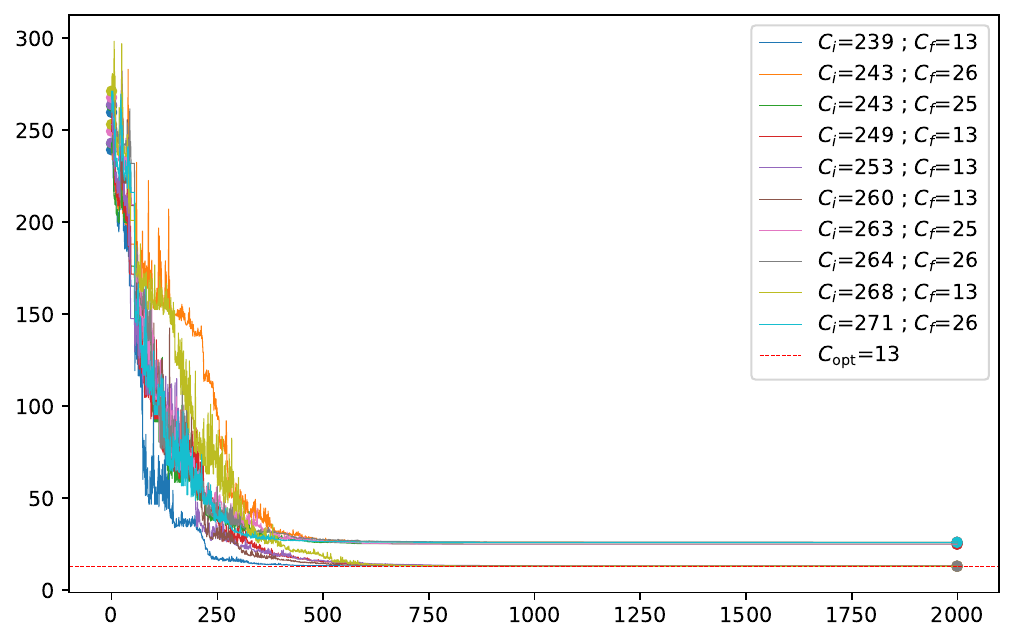}
        \vspace{0.5em}
        \textbf{(b)}
    \end{minipage}

    \caption{VQE optimization processes starting from (a) the 10 best MUBs (b) 10 random initial points. \: Initial and final costs are denoted by $C_i$ and $C_f$ respectively. Optimal cost ($C_{opt}$) appears for comparison.}
    \label{fig:MUBsRand}
\end{figure}

\begin{figure}[h!]
    \centering
    \includegraphics[width=0.7\linewidth]{{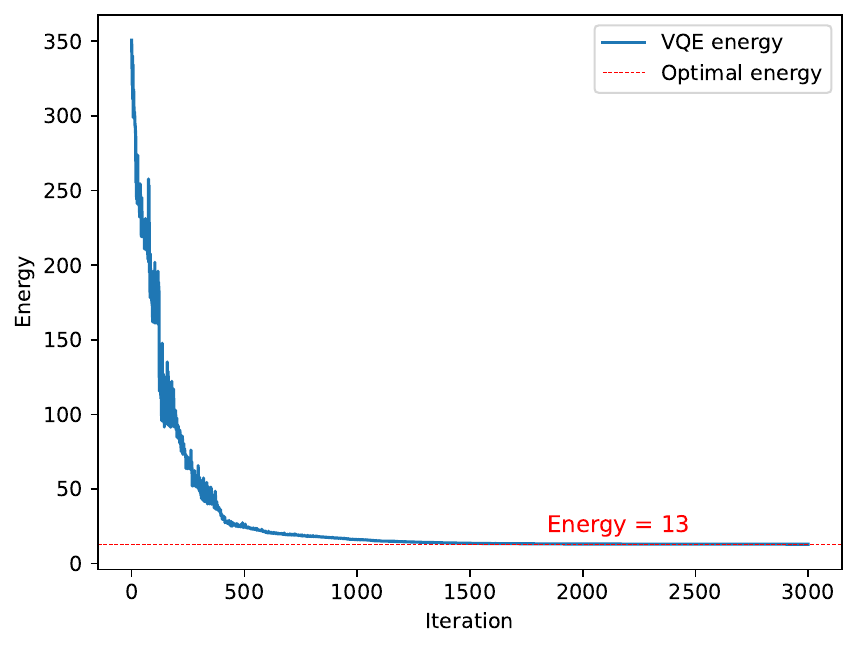}}
    \caption{VQE simulation results for the 4-node TSP instance from \Fig{fig:Graph_K1}. Convergence for the known minimal cost (dashed red line) was achieved.}
    \label{fig:VQE}
\end{figure}

Here, we experimented with a computationally faster approach than starting from all the MUB states of the partial-DQES. Instead, we considered starting from a small number $k$ of lowest energy MUB states in the MUB landscape. We are of course not promised that it is always a good thing to do, as maybe a 'bad' point is the only one that bypasses a barren plateau. Nevertheless, the intuition behind starting from the best points is straightforward: It is plausible to hope that starting from low energy states which are (in some sense) closer to the ground-state can improve convergence. An additional benefit is that the number of VQE runs is now $k=O(1)$ instead of $O(n)$.
\\
For the TSP defined in \Fig{fig:Graph_K1}, we ran VQE starting from the best $k=10$ MUB states in the landscape obtained in previous section for the same instance. As a comparison, we also ran VQE with random initializations; an alternative approach used to avoid barren plateaus and local minima (see \cite{ref:kim2022quantum} and \cite{ref:miyahara2022ansatz}), for example). We started from $k=10$ different random points. The results are shown in \Fig{fig:MUBsRand}. 
From \Fig{fig:MUBsRand} we can tell that 6 out of the 10 VQE runs starting from best MUB states converged, compared to 5 out of the 10 runs starting from random points. More comparisons (and on other problems) are required to deduce a meaningful advantage to either of the methods, but we can see that at least for this TSP instance, both methods are useful in avoiding local minima or barren plateaus. 
\\
The results show an additional feature according to \Fig{fig:MUBsRand}. We see that on average, fewer iterations are required for convergence when starting from the best MUB states, compared to starting from random points. This agrees with our intuition, as the best landscape points are closer to the ground-state. This can be another potential advantage of using the method of starting from the best points in the landscape. 
\\
Following the standard practice, we also include a VQE run for starting from $\ket 0^{\otimes n}$. This result shows a successful convergence and is shown in \Fig{fig:VQE}, with the convergence process seen more clearly on a single curve graph. 
All our VQE results were obtained by running noiseless simulators via Qiskit \texttt{Python} package, using COBYLA as the optimizer and the Efficient-$SU(2)$ circuit (by default), as the ansatz.  
\section{Discussion and Further Work}

\label{sec:Discussion}
In this work we started by clarifying with details the Ising model for the TSP, along with qubit efficiency and we provided various important notes.  

We then applied the partial-DQES on small instances of the TSP, a prominent NP-complete problem. We also ran VQE simulations for the same instance. We observed how identifying symmetries in the problem allows for a reduction of the required Hilbert space dimension. Here a constant reduction was achieved. Such a reduction can nevertheless be relevant in the NISQ era, or even in classical simulations, as it allows us to simulate slightly larger instances.

This work is an initial exploration of the DQES method, specifically for the TSP. As such, there are several directions in which we can further investigate:
\begin{itemize}
    \item \textbf{Further comparisons between the DQES method and other methods:} This was already demonstrated here with the standard VQE, but we aim to simulate larger instances of the problem and examine specifically how DQES can help with trainability issues such as barren plateaus and local minima. We can compare with other methods that claim to be effective in avoiding barren plateaus, such as \cite{ref:grant2019initialization}. Comparison to QAOA, or even adapting DQES to QAOA, is a potential direction as well. In addition, we hope to perform experiments with noisy simulators or on real noisy devices. 
    \item \textbf{Formulating concrete heuristics for DQES-based SAT- or LH- solvers:} The method we used in our simulations is partial-DQES for $k=2$ qubits. According to \cite{ref:yu2023MUBsCirc}, building the quantum circuit that generates MUBs over $k$ qubits can be done efficiently, and even classically, as the gates required are all Clifford gates. The caveat is that the number of MUB states is exponential in $k$, so the whole procedure of trying all MUB states is not efficient. \\
    Despite the inefficiency of naively performing full-DQES on large systems, there are many possibilities worth exploring: (1) There is freedom in how to initialize the other qubits, for example by random initialization in the computational basis, in the whole Hilbert space, or initialization that exploits symmetries or properties of the problem of interest. (2) Knowing how to generate MUB states, we can perform a full DQES on small systems and see if relevant structures (in the energy landscape, for example) can be identified that may help in tackling larger problems. (3) Beyond initialization, we are interested in developing an optimizer that searches near MUB points, and incorporates problem-specific heuristics as well.  
    \item \textbf{Develop and incorporate more qubit efficiency techniques:} By identifying further symmetries or using encoding tricks, such as the logarithmic encoding discussed in \Sec{sec:Qubit-eff2} and \cite{ref:Lucas_2014}, we may be able to further reduce the number of qubits required for the TSP, or apply such techniques to other problems. 
    \item \textbf{Other NP-complete problems:}  
    The Traveling Salesman Problem (TSP) admits many generalizations, such as the Vehicle Routing Problem (VRP), which in turn has numerous variants. These generalizations are especially relevant for real-world applications, for instance in logistics and manufacturing. At the same time, TSP itself, perhaps surprisingly, already plays a role in the context of DNA sequencing \cite{ref:sarkar2021DNA}, a problem of immense importance and one that we wish to study in more detail as well. We believe that considering both generalizations of TSP (such as VRP) and specializations of TSP (such as DNA sequencing) can be beneficial for algorithm design: in the former case, by exposing broader structural principles and symmetries that persist across related problems; in the latter, by exploiting domain-specific structure that may allow for more efficient algorithms. 
\end{itemize}

Ultimately, our exploration of DQES for TSP highlights the potential of symmetry-based and structure-aware approaches in quantum optimization. While many open questions remain, especially regarding scalability and efficiency, the flexibility of the DQES framework—ranging from problem-specific heuristics to generalizations across different NP-complete problems—suggests a rich space for further development. We hope that progress in this direction will help clarify the potential advantages of DQES relative to existing variational methods, and contribute to a deeper understanding of how quantum algorithms can exploit problem structure. We emphasize that DQES for the VRP and for analyzing molecules ground
state anaergy is further dealt with by \cite{ref:GMMY, ref:GMR}.

%
%
%
\section*{Acknowledgments}
We thank Lev Yohananov for
fruitful discussions. T.M and O.G thank the
Quantum Computing Consortium of Israel Innovation Authority for financial
support. This research project was partially supported by the Helen Diller
Quantum Center at the Technion.
\bibliographystyle{ieeetr}
\bibliography{KGB}
\end{document}